# Significancia estadística del exceso de muertes en Chile durante pandemia COVID-19

Alejandra Matus[1], Marcelo Matus[2], Felipe Cabello[3] y Erick Sierra[4]

INTRODUCCIÓN

El estudio de la mortalidad y el cálculo del exceso de muertes es una herramienta fundamental para el entendimiento del efecto de las epidemias/pandemias [1] y de los ciclos estacionales de enfermedades endémicas/epidémicas como la influenza [2], que resulta de gran importancia en tiempos de crisis, según la OMS y diferentes autores [3, 4, 5].

El exceso de muertes cuantifica la diferencia entre las muertes reportadas y el número de muertes esperadas en un periodo determinado. El análisis y el monitoreo del exceso de muertes ayuda a la toma de decisiones de políticas de salud informadas y a entender el efecto social y económico de una epidemia/pandemia [6].

El indicador es utilizado regularmente por agencias o proyectos públicos, como el EuroMOMO[1], pero se ha vuelto de especial interés en el caso de la actual pandemia (COVID-19), porque ha permitido obtener información sobre la evolución de la mortalidad más allá de la información y cifras oficiales de muertes por esta enfermedad, que pueden estar subestimadas ya sea por diagnóstico equivocado y/o por sub-reporte de estas muertes [7].

El exceso se calcula comparando las muertes reportadas durante un periodo sobre un "benchmark" o valor de referencia del número esperado de muertes, construido en base a información histórica en tiempos normales para igual periodo. Habitualmente el valor de referencia se construye como el promedio (valor esperado) de las muertes reportadas en los últimos 3 a 5 años para la población completa o considerando sub-rangos etarios y regionales [8, 7, 9, 10].

EuroMOMO, por ejemplo, utiliza un algoritmo [11] que analiza los datos de los últimos 3 y 5 años, por rangos etarios y sub-regiones, en periodos de semanas, para 24 estados europeos participantes desde el 2008.

---

[1] EuroMOMO is a European mortality monitoring entity, aiming to detect and measure excess deaths related to seasonal influenza, pandemics and other public health threats. Official national mortality statistics are provided weekly from the 24 European countries and regions in the EuroMOMO collaborative network, supported by the European Centre for Disease Prevention and Control (ECDC) and the World Health Organization (WHO). https://www.euromomo.eu/

El CDC (Centers for Disease Control and Prevention) de Estados Unidos, que calcula el exceso de muertes por COVID-19 desde el 2 de febrero de este año [12], construye dos valores de referencias con la información de la mortalidad del 2015 al 2019: el promedio y el limite superior del intervalo de confianza de 95%. Con esto se calcula el exceso de muertes por semana y por jurisdicción [13].

Finalmente, varias publicaciones internacionales han desarrollado sistemas de monitoreo de exceso de muertes de países y regiones alrededor del mundo:

1. Financial times [14], que usa el promedio histórico de 2 a 5 años como referencia para el periodo de análisis desde el inicio del contagio por país,
2. The Economist [15], que usa el promedio del 2015 al 2019 para periodos semanales,
3. The New York Times [16], que partió utilizando promedios históricos, hoy utiliza un método de ajuste lineal[2] y suavización para países con cinco años de información y promedio para regiones con solo tres años de información.

Otras publicaciones internacionales, como The Washington Post [17] y The Guardian [18], han reportado sobre el indicador de exceso de muertes y la importancia de su inclusión en los análisis de la evolución de la pandemia y la toma de decisiones públicas.

Si bien el exceso de muertes es una herramienta relevante, no debe entenderse como un reemplazo a la contabilización y clasificación correcta de las muertes durante una pandemia. No se debe esperar tampoco que coincida con el número de muertes oficiales reportados por causa COVID-19, por ejemplo, ni en el mejor el sistema de salud del mundo. Esto porque el indicador, por definición, incluirá las muertes de todas las causas.

Este indicador debe usarse como una herramienta más, que ayuda al entendimiento del impacto de la pandemia, y es su evolución, aumentos o disminuciones, lo que puede de mejor manera servir para la toma de decisiones más informada y pertinente de pate de la ciudadanía y las autoridades.

Exponemos aquí, textualmente, la petición de diversos investigadores, publicada en Lancet, para que los países generen bases nacionales de excesos de muertes en línea, con tabulaciones por sexo y grupos de edades, a la luz de la magnitud de la pandemia COVID-19 [19]:

*"Weekly excess deaths could provide the most objective and comparable way of assessing the scale of the pandemic and formulating lessons to be learned. We therefore urge all national authorities who can collate counts of weekly deaths to expedite the publication of these data and place them in the public domain. The dissemination of this information should be done*

---

[2] El método actual de corrección es propio del NYT y no ha sido validado en una publicación revisada por pares. La estimación y corrección lineal de series de tiempo en presencia de una distribución subyacente ($y_i = a + b\, t_i + n(t_i)$), introduce errores en la estimación de la tasa de crecimiento, que pueden llevar a sobre o subestimaciones del exceso de muertes.

*within 3–4 weeks of the period of observation. At a minimum, tabulations by sex and 5-year age groups are essential."*

SITUACIÓN EN CHILE

Mientras el exceso de muertes ha sido aceptado por la comunidad internacional como una herramienta adicional para observar y entender la evolución y el impacto de la pandemia, particularmente, su posible costo en vidas, tanto directas como indirectas, en Chile no hay ninguna agencia nacional calculando este indicador, hasta donde los autores de este trabajo conocen.

Es más, se ha considerado que el indicador no es relevante para el país puesto que la probabilidad de tener muertes no reportadas por COVID-19 ha sido declarada a priori "prácticamente nula" [20].

Sin embargo, en Chile, el sistema actual de reporte de muertes es complejo, e incluye a diversas agencias públicas y agentes privados, que no necesariamente actúan efectiva y coordinadamente, pero sí bajo condiciones de demanda sistémica nunca vista.

Esto no ocurre en un laboratorio, sino en un sistema sanitario, con sus prácticas consuetudinarias, que replican la desigualdad estructural, que ha sido suficientemente analizada y reconocida desde distintas disciplinas. Sin contar esa inmensa proporción de la realidad, que incluso ha escapado a la mirada de los sistemas de salud más robustos y sociedades con mejor protección social que la nuestra, que es el de las muertes domiciliarias, que, según datos del Registro Civil para muertes respiratorias del mes de marzo 2020, pueden representan un 46% del total [21]. De estas muertes domiciliarias, la mayoría de las que sean por causa COVID-19 no calificarán para ser incluidas en las cifras oficiales.

La Tabla 1, de análisis de exceso de muertes, muestra el número de fallecidos por regiones del 2015 al 2020, entre el 2/3/2020 y el 7/6/2020, el cálculo de exceso de muertes utilizando: el valor de referencia del promedio de los últimos cinco años (PUCA) y el límite superior del intervalo de confianza de 95% obtenido para el mismo PUCA, en base a la desviación estándar (DUCA) y dos distribuciones, la Normal y la Tstudent. Los datos se obtuvieron del Registro Civil a la fecha del 7/6/2020 [22].

| Análisis de exceso de muertes por regiones acumulado del 2/3/2020 al 7/6/2020 | | | | | | | | | | | | Criterio IC 95% Normal | | | Criterio IC 95% Tstudent | | |
|---|---|---|---|---|---|---|---|---|---|---|---|---|---|---|---|---|---|
| Regiones | 2015 | 2016 | 2017 | 2018 | 2019 | 2020 | PUCA | DUCA | Exceso muertes | Exceso / PUCA P-score | Exceso / DUCA Z-score | PUCA+ Lim Sup IC 95% | Exceso+ muertes | Exceso+ / PUCA+ P-score | PUCA+ Lim Sup IC 95% | Exceso+ muertes | Exceso+ / PUCA+ P-score |
| Metropolitana de Santiago | 10.081 | 10.201 | 10.512 | 10.456 | 10.994 | 13.482 | 10.449 | 353 | 3.033 | 29,0% | 8,6 | 10.758 | 2.724 | 25,3% | 10.887 | 2.595 | 23,8% |
| Tarapacá | 384 | 389 | 362 | 353 | 386 | 465 | 375 | 16 | 90 | 24,1% | 5,6 | 389 | 76 | 19,5% | 395 | 70 | 17,7% |
| Antofagasta | 784 | 816 | 791 | 820 | 792 | 853 | 801 | 16 | 52 | 6,5% | 3,2 | 815 | 38 | 4,7% | 821 | 32 | 3,9% |
| Valparaíso | 3.088 | 3.161 | 3.194 | 3.248 | 3.491 | 3.641 | 3.236 | 154 | 405 | 12,5% | 2,6 | 3.371 | 270 | 8,0% | 3.427 | 214 | 6,2% |
| Biobío y Ñuble | 3.224 | 3.192 | 3.421 | 3.418 | 3.465 | 3.640 | 3.344 | 126 | 296 | 8,9% | 2,3 | 3.454 | 186 | 5,4% | 3.501 | 139 | 4,0% |
| Magallanes | 255 | 263 | 300 | 294 | 300 | 333 | 282 | 22 | 51 | 17,9% | 2,3 | 301 | 32 | 10,5% | 309 | 24 | 7,7% |
| Arica y Parinacota | 367 | 314 | 341 | 323 | 326 | 362 | 334 | 21 | 28 | 8,3% | 1,3 | 352 | 10 | 2,7% | 360 | 2 | 0,6% |
| Los Ríos | 671 | 652 | 687 | 644 | 718 | 714 | 674 | 30 | 40 | 5,9% | 1,3 | 700 | 14 | 2,0% | 711 | 3 | 0,4% |
| Atacama | 422 | 380 | 373 | 371 | 373 | 407 | 384 | 22 | 23 | 6,0% | 1,1 | 403 | 4 | 1,1% | 411 | - | |
| La Araucanía | 1.653 | 1.693 | 1.747 | 1.654 | 1.863 | 1.813 | 1.722 | 88 | 91 | 5,3% | 1,0 | 1.799 | 14 | 0,8% | 1.831 | - | |
| O'Higgins | 1.363 | 1.342 | 1.339 | 1.431 | 1.486 | 1.445 | 1.392 | 64 | 53 | 3,8% | 0,8 | 1.449 | - | | 1.472 | - | |
| Coquimbo | 1.067 | 1.026 | 1.036 | 1.163 | 1.206 | 1.163 | 1.100 | 80 | 63 | 5,8% | 0,8 | 1.170 | - | | 1.199 | - | |
| Aysén | 110 | 119 | 154 | 114 | 134 | 140 | 126 | 18 | 14 | 10,9% | 0,8 | 142 | - | | 149 | - | |
| Maule | 1.655 | 1.557 | 1.658 | 1.756 | 1.750 | 1.724 | 1.675 | 82 | 49 | 2,9% | 0,6 | 1.747 | - | | 1.777 | - | |
| Los Lagos | 1.338 | 1.327 | 1.348 | 1.390 | 1.349 | 1.335 | 1.350 | 24 | (15) | -1,1% | -0,6 | 1.371 | - | | 1.380 | - | |
| **Total Regiones** | 26.462 | 26.432 | 27.263 | 27.435 | 28.633 | 31.517 | 27.245 | | 4.272 | 15,7% | | 28.222 | 3.367 | 11,9% | 28.629 | 3.079 | 10,8% |
| **Chile** | 26.462 | 26.432 | 27.263 | 27.435 | 28.633 | 31.517 | 27.245 | 900 | **4.272** | 15,7% | 4,7 | 28.034 | **3.483** | 12,8% | 28.362 | **3.155** | 11,1% |

PUCA: Promedio últimos cinco años (2015-2019)  
DUCA: Desviación estándar últimos cinco años (2015-2019)  
Exceso muertes: diferencia entre muertes de 2020 y PUCA  
Fuente: Registro Civil https://estadisticas.sed.srcei.cl/defreg, datos al 7/6/2020  
Fecha de inicio 2/3/2020 corresponde al primer contagiago de COVID confirmado  
Fecha de término 7/6/2020 incluye 14 semanas completas  
Se destacan valores de Exceso/DUCA con significación estadística (>2 StdDev)  
Exceso+ y PUCA+ se construyen en base a lím. sup. del IC 95%, según CDC-USA con distr. Normal y Tstudent

Tabla 1: Análisis de exceso de muertes por regiones desde el 3/2/2020 al 7/6/2020.

Según se puede observar, el comportamiento del exceso de muertes es fuertemente dependiente de la región considerada. La significancia estadística para las regiones Metropolitana y de Tarapacá es notoria, con valores de Z-score de 8,6 y 5,6. También se observan otras cuatro regiones con valores de Z-score por sobre 2, y con un valor total para el país de 4,7.

En la tabla se muestra que el exceso de muertes para los valores de referencia utilizados: totalizando 4.272 para es caso de PUCA, que corresponde a un 15,7% de incremento (P-score); 3.483 para el caso de usar los límites superiores del IC de 95% con distribución Normal, que corresponde a 12,8%; y finalmente de 3.155.

El exceso de muertes igualmente puede ser calculado en periodos semanales, lo que permite analizar su evolución en el tiempo.

La Figura 1 muestra la evolución del exceso de muertes para el país considerado como una sola región. Aquí el "Exceso" fue calculado similar a lo realizado en la Tabla 1, pero usando periodos semanales. Como se observa, los cálculos semanales y acumulados solo coinciden cuando se compara el método de PUCA, puesto que el cálculo de los intervalos de confianza no es lineal.

Estas curvas son calculadas directamente desde la información publicada por el portal de datos Registro Civil [22], a través de un script de acceso público [23], escrito en base al portal del Ministerio de Educación [24].

Favor de notar la inflexión pronunciada observada para las curvas de Exceso y Exceso+ para la semana del 3 de mayo, que puede verse reflejada en la inflexión de muertes por COVID-19 que comienza la semana del 17 de mayo.

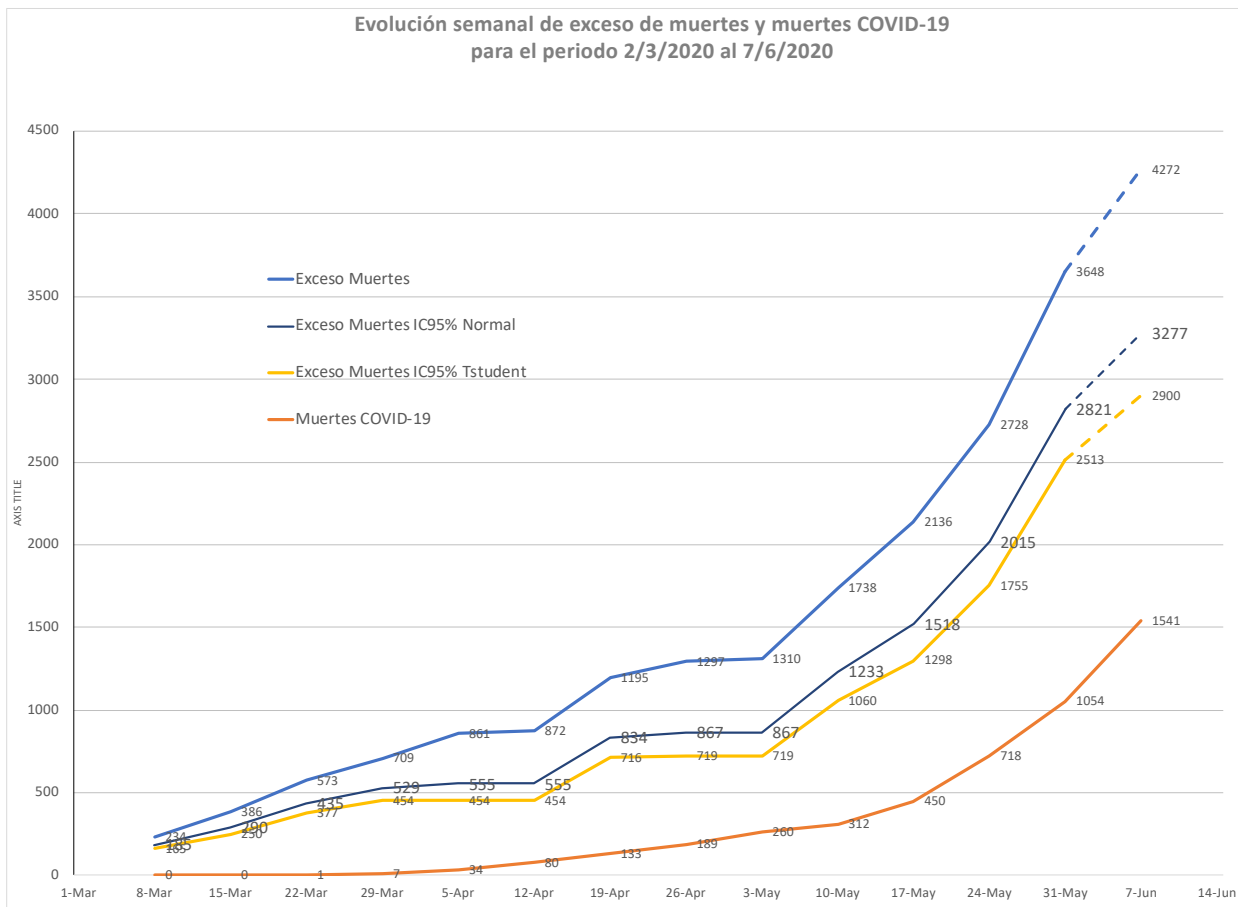

Figura 1: Evolución del exceso de muertes semanal de Chile calculado semanalmente desde 3/2/2020 al 31/5/2020, en base a defunciones inscritas del Registro Civil disponibles al 1/6/2020. Se muestra la curva de "Exceso" (azul), que utiliza como valor de referencia el promedio de las defunciones inscritas de los últimos cinco años (PUCA), la de "Exceso IC95% Normal"(gris), que considera el límite superior del intervalo de confianza (95%) con distribución Normal y de "Exceso IC95% Tstudent" (amarilla=, que considera el límite superior del intervalo de confianza (95%) con distribución Tstudent. Para referencia se incluye el número de muertes COVID-19 reportados a la fecha (naranja). Los últimos tramos del cálculo de excesos está con línea punteada dado que el registro solo ha publicado parcialmente los valores de la última semana.

## Críticas al significado estadístico del exceso de muertes

Desde que comenzamos a calcular y publicar el exceso de muertes de Chile, con gran dificultad y usando, en el primer intento, datos del registro civil obtenidos en fuente reservada, puesto que los datos no eran públicos, surgieron varias voces argumentando que la metodología aplicada mundialmente no podía utilizarse en Chile. Se han dado entre otras razones el tamaño del país y la manera en que el registro Civil está entregando la información, que corresponde a la fecha de la inscripción (que puede retrasarse, especialmente los fines de semana) y no a la de la fecha de muerte efectiva. Que en nuestro caso debía incorporarse normalización por población, correcciones por gran aumento de inmigrantes, el número de lunes incorporados en el periodo de análisis, o el aumento porcentual de las muertes que ocurre cada año.

Luego, junto a la crítica de si aplicaba o no la metodología en Chile, comenzaron a desarrollarse diferentes mecanismos de corrección y cálculos alternativos que demostraban, a juicio de los proponentes, si el cambio era o no significativo, con preferencia de la última opción.

Al respecto, los proponentes de las metodologías de corrección emiten sus juicios respecto de la significancia de los datos una vez corregidos, lo cual hace que la conclusión esté directamente correlacionada con el mecanismo desarrollado, y no necesariamente con los datos observados directamente.

De los diferentes argumentos y metodologías que introducen una corrección a priori, observamos al menos dos familias de propuestas:

1. Que el exceso de muertes del 2020, el 2019, PUCA, o entre otros años es similar. Ergo, las variaciones observadas no son relevantes.
2. Que el exceso de muertes 2020 sí es observable, pero se puede explicar por el crecimiento vegetativo del número de muertes observada en las estadísticas de largo plazo, que estaría a su vez relacionada con el aumento de la población.

**Hipótesis 1: Exceso de muertes para 2020 es similar a otros periodos y no es relevante**

Esta hipótesis se puede testear estadísticamente haciendo el análisis de la muestra y con la ayuda de la desviación estándar, que nos permite entender cuán variable es.

Una forma de establecer si el número de muertes del 2020 está dentro de los valores que normalmente podrían ocurrir sin pandemia, es estimar, con un nivel de confidencia deseado, el rango normal esperado de posibles valores en función de la desviación estándar observada en los años anteriores.

Si quisiéramos usar, por ejemplo, un nivel de confidencia de 95%, podemos afirmar que si el valor de muertes de 2020 supera el PUCA por sobre 1.96 veces la desviación estándar de los últimos años cinco años, la diferencia es significativa.

Así, determinar si el exceso de muertes es significativo puede ser testeado, antes de implementar una corrección a priori, calculando la razón entre el exceso de muertes del 2020 y la desviación estándar de los mismos cinco años.

En la Tabla 1, la columna "Exceso/DUCA" calcula esta razón, también conocido como el Z-score.

**Hipótesis 2: Exceso de muertes para 2020 se explica por crecimiento lineal de muertes/población**

Si el exceso de muertes se pudiese explicar por un crecimiento lineal, es posible formular una prueba estadístico simple en función de la misma desviación estándar de la muestra, antes de aplicar cualquier corrección.

Supongamos por un momento que la muestra de los seis años crece con una tasa lineal de crecimiento fija, entonces si el exceso de muertes es superior a 1.9 veces la desviación estándar[3], esto representaría un valor mayor que el proyectado y debe entenderse como significativo.

**Corrección del exceso de muertes**

De la discusión anterior, es fácil verse tentado a proponer un "método de corrección" para calcular el verdadero valor de exceso de muertes, suponiendo, por ejemplo, que el proceso corresponde a un crecimiento lineal, o utilizar algún otro método ad-hoc de corrección tan simplificado o complejo como se desee.

Sin embargo, esto es poco recomendable en atención a que no se puede asumir que el año 2020, por la interferencia de la pandemia, se comportará en forma "normal". Unas causas habituales de deceso pueden disminuir y otras aumentar. El Ministerio de Salud ha informado que las enfermedades respiratorias, que normalmente comienzan a circular en marzo, no han estado presentes en forma significativa este año [25].

Igualmente, como se observa en la Tabla 2, es difícil aplicar un criterio uniforme cuando es posible observar que el comportamiento no es igual para todas las regiones. No se puede decir, por ejemplo, que las defunciones crecen linealmente en todas las regiones a la misma tasa, o que la desviación estándar es similar para todas.

Es igualmente importante señalar que, para marzo, el único mes en el que tuvimos a la vista las causas de muerte, la distribución etaria también se comportó de un modo significativamente distinto al promedio de los años previos. Los decesos de jóvenes y niños disminuyeron, y en cambio aumentaron significativamente la de mayores de 50, aumentando su proporción a mayor edad, según se puede observar en el resumen de la Tabla 2.

---

3 3 La serie lineal para los seis años tendría la siguiente estructura:

| Año | 2015 | 2016 | 2017 | 2018 | 2019 | 2020 |
|---|---|---|---|---|---|---|
| Muertes | $M_0 - 2r$ | $M_0 - r$ | $M_0$ | $M_0 + r$ | $M_0 + 2r$ | $M_0 + 3r$ |

Esta estructura es determinística, en el sentido que se puede proyectar el valor esperado de muertes para el 2020 en forma exacta y podemos calcular el promedio y la desviación estándar de los primeros cinco años, resultando:

$$\sigma = \sqrt{\tfrac{5}{2}}\, r, \ \ PUCA = M_0$$

Usando la relación anterior, se puede obtener el exceso proyectado para 2020 de 3$r$, donde $3r = 3\sqrt{\tfrac{2}{5}}\,\sigma$, o bien 1.90 $\sigma$.

| Analisis de exceso de muertes marzo 2020 por rango de edad | | | | | | | | | | |
|---|---|---|---|---|---|---|---|---|---|---|
| Rango Edad | 2015 | 2016 | 2017 | 2018 | 2019 | 2020 | PUCA | DUCA | Exceso | Exceso / DUCA | Exceso / PUCA |
| de 0 a 19 | 233 | 230 | 225 | 184 | 204 | 168 | 215 | 21 | -47 | -2,3 | -22% |
| de 20 a 39 | 424 | 359 | 363 | 371 | 382 | 406 | 380 | 26 | 26 | 1,0 | 7% |
| de 40 a 59 | 1228 | 1229 | 1244 | 1134 | 1065 | 1208 | 1180 | 78 | 28 | 0,4 | 2% |
| de 60 a 79 | 3037 | 3091 | 2966 | 3113 | 3134 | 3471 | 3068 | 68 | 403 | 6,0 | 13% |
| de 80 a 109 | 3102 | 2949 | 3044 | 3100 | 3045 | 3509 | 3048 | 62 | 461 | 7,4 | 15% |
| Total | 8024 | 7858 | 7842 | 7902 | 7830 | 8762 | 7891 | 79 | 871 | 11,0 | 11% |

Tabla 2: Análisis de exceso de muertos para marzo 2020 por rango de edad a partir de datos de inscripción de defunciones. Fuente: reservada.

La corrección por crecimiento vegetativo no explica este comportamiento. Tampoco la incorporación de población migrante (en la que majaderamente algunos han intentado encontrar la explicación), pues, por el contrario, las cifras oficiales de población migrante en Chile dan cuenta que el 74,3% se ubica en el tramo etario inferior a 39 años [26], por lo cual se puede afirmar que la población migrante en Chile es mayoritariamente una población joven.

**Cuestionamiento a las fuentes**

En su réplica a un artículo que publicamos en El Mostrador, junto al profesor Felipe Cabello, la ex directora del Departamento de Estadísticas del Ministerio de Salud (DEIS), Dra. Danuta Rajs, cuestiona que se use como fuente las inscripciones anotadas por el Registro Civil y sostiene, como ha dicho el Ministro de Salud, que ese es un registro incompleto (pues está continuamente agregando información de muertes ocurridas, por ejemplo, fuera de Chile) y que las causas son precisadas por el DEIS y el INE en un proceso posterior. También indica que la corrección demográfica es necesaria. Sobre la base de los datos del DEIS/INE, oficiales y publicados hasta 2017, y datos que reconoce como "provisorios" para los años 2018-2020, la profesora Rajs sostiene que el exceso de muertes registradas en el período es normal.

Los cálculos que se han hecho en otros países usan también los de instituciones equivalentes al Registro Civil en Chile, a pesar del retraso que todos tienen en incorporar fallecimientos cuya inscripción demora por razones tanto extraordinarias, como excepcionales. El margen de error de esa fuente, puesto que es la misma utilizada para el cálculo completo en nuestro caso, se presume similar para todos los años y, respecto de 2020, donde debería existir un mayor número de muertes no inscritas, la posibilidad es que el número crezca (y no disminuya) en comparación con los años anteriores. No nos es posible usar el cálculo provisorio del DEIS, pues la solicitud por Transparencia que hicimos en abril aún está pendiente de respuesta y porque, dada la naturaleza del trabajo que se hace sobre ellos, el margen de error es mayor en comparación con los años de datos oficiales ya publicados hasta 2017. Es decir, nos parece más incierto usar como fuente los datos provisorios del DEIS, que según reconoció el INE en respuesta a otra solicitud de Transparencia que le hicimos, demora al menos dos años en estabilizar los números de un período en particular.

Más allá del análisis estadístico, complementa nuestro análisis la evidencia obtenida con métodos periodísticos en terreno: casos concretos de personas cuyos certificados de defunción no registran COVID-19 como causa de muerte, aunque los fallecidos dieron positivo al test PCR; testimonios de dueños de funerarias que relatan haber sido informados de casos positivos, aunque no se anotaban así en los certificados oficiales; fallecimientos domiciliarios en que, a pesar de la sospecha de COVID-19, las personas no fueron testeadas; estadísticas de tres grandes cementerios (Cementerio General, Cementerio Playa Ancha y Cementerio Metropolitano) que detectan los mismos fenómenos: aumentos significativos de sepulturas en el período de pandemia, aumento de defunciones con causas respiratorias y en los que, apenas una fracción, se anotan como casos COVID-19.

En resumen, los procesos subyacentes al comportamiento del exceso de mortalidad del año 2020 son más complejos que un crecimiento lineal, y cualquier corrección requiere más datos desagregados, como causas de muerte, edad, los cuales aún no están disponibles. Es probable que, en el futuro, con datos más precisos, se pueda hacer análisis más detallados sobre el exceso de muertes y sus causas durante la pandemia.

Sin embargo, consideramos ineludible analizar el comportamiento de este indicador para contrarrestar, por ejemplo, la tasa de letalidad, pues la autoridad puede contar así, con nuestros análisis y los alternativos que se han propuesto, un faro de luz para guiarse en estos tiempos cada vez más oscuros que se avecinan y desarrollar más y mejores políticas públicas destinadas a proteger la salud de los chilenos y chilenas.

Conclusiones

Mientras se escribía este artículo, la postura del Ministerio de salud ha cambiado. Cuando publicamos los primeros cálculos, el ministro de salud descartó tajantemente un aumento atípico en la mortalidad en Chile, declarando que "*A veces se hacen matemáticas fáciles y conclusiones apresuradas*" [27].

Hoy, el ministro declara *"la necesidad de reconocer, según recomendaciones de la OMS, tanto en marzo como en abril, a asignar como posibles casos de muertes, asociadas a Covid-19, un número de 653 personas fallecidas"* [28].

Este es un gesto importante de las autoridades de salud, sin embargo, aún utilizando el método más conservador presentado en este estudio, existe un exceso de muertes que requiere entendimiento y explicación, para una mejor toma de decisiones y reducción de los efectos de la pandemia.

Para ello, esperamos que tanto el Ministerio de Saludo como los otros organismos del estado, DEIS, Registro Civil y el Instituto de Salud Pública, dispongan de la información que sea útil para que la comunidad civil y científica puedan analizar los datos de manera independiente.

Autores

[1] Alejandra Matus, Periodista, MPA/Harvard University, profesora asociada Universidad Diego Portales.

[2] Marcelo Matus, PhD en Ingeniería Eléctrica y Computacional con mención en matemáticas aplicadas, University of Arizona, investigador en el Centro de Energía, Facultad de Ciencias Físicas y Matemáticas de la Universidad de Chile.

[3] Felipe Cabellos, MD, Professor Department of Microbiology and Immunology, New York Medical College.

[4] Erick Sierra, Ingeniero Civil Eléctrico, investigador en el Centro de Energía, Facultad de Ciencias Físicas y Matemáticas de la Universidad de Chile.

Bibliografía

[1] J. M. Gran, J. M. Gran, O. Kacelnik, A. M. Grjibovski, P. Aavitsland, B. G. Iversen y B. G. Iversen, «Counting pandemic deaths: comparing reported numbers of deaths from influenza A(H1N1)pdm09 with estimated excess mortality,» *Influenza and Other Respiratory Viruses,* vol. 7, nº 6, pp. 1370-1379, 2013.

[2] M. Park, P. Wu, E. Goldstein, W. J. Kim y B. J. Cowling, «Influenza-Associated Excess Mortality in South Korea,» *American Journal of Preventive Medicine,* vol. 50, nº 4, 2016.

[3] M. R. E. O.-O. a. J. H. Hannah Ritchie, «Excess mortality from the Coronavirus pandemic (COVID-19),» Our Workd in Data, 2020. [En línea]. Available: https://ourworldindata.org/excess-mortality-covid.

[4] F. Checchi y L. Roberts, «Interpreting and using mortality data in humanitarian emergencies A primer for non-epidemiologists,» *Humanitarian Practice Network,* nº 52, September 2005.

[5] Wolrd Health Organization, «Defininition: emergencies,» WHO, 2020. [En línea]. Available: https://www.who.int/hac/about/definitions/en/.

[6] J. Aron y J. Muellbauer, «Measuring excess mortality: England is the European outlier in the Covid-19 pandemic,» 18 May 2020. [En línea]. Available: https://voxeu.org/article/excess-mortality-england-european-outlier-covid-19-pandemic.

[7] J. Aron y J. Muellbauer, Measuring excess mortality: the case of England during the Covid-19 Pandemic, Oxford: INET Oxford Working Paper No. 2020-11, 2020.

[8] K. M. a. M. Ebeling, «EXCESS MORTALITY FROM COVID-19. WEEKLY EXCESS DEATH RATES BY AGE AND SEX FOR SWEDEN.,» *medRxiv ,* vol. Preprint, 2020.

[9] C. Gabriele y S. Garcia-Mandicó, «COVID-19 in Italy: An analysis of death registry data,» 22 April 2020. [En línea]. Available: https://voxeu.org/article/covid-19-italy-analysis-death-registry-data.

[10] New York City Department of Health and Mental Hygiene (DOHMH) COVID-19 Response Team, «Preliminary Estimate of Excess Mortality During the COVID-19 Outbreak — New York City, March 11–May 2, 2020.,» *MMWR Morb Mortal Wkly Rep,* vol. 69, nº Preprint, p. 603–605, 2 May 2020.

[11] B. Gergonne, European algorithm for a common monitoring of mortality across Europe, EuroMOMO Work Package 7, 2011.

[12] GlobalData Healthcare, «US Centers for Disease Control and Prevention (CDC) and the National Center for Health Statistics (NCHS),» Pharmaceutical Technology, 21 May 2020. [En línea]. Available: https://www.pharmaceutical-technology.com/comment/excess-deaths-pandemic-covid-19-fatalities/.

[13] CDC, «Excess Deaths Associated with COVID-19,» Centers for Disease Control and Prevention, 29 May 2020. [En línea]. Available: https://www.cdc.gov/nchs/nvss/vsrr/covid19/excess_deaths.htm.

[14] Financial TImes, «Global coronavirus death toll could be 60% higher than reported | Free to read,» Financial Times, 2020. [En línea]. Available: https://www.ft.com/content/6bd88b7d-3386-4543-b2e9-0d5c6fac846c.

[15] The Economist, «Tracking covid-19 excess deaths across countries,» 2020. [En línea]. Available: https://www.economist.com/graphic-detail/2020/04/16/tracking-covid-19-excess-deaths-across-countries.

[16] NYT, «87,000 Missing Deaths: Tracking the True Toll of the Coronavirus Outbreak,» The New York Times, May 2020. [En línea]. Available: https://www.nytimes.com/interactive/2020/04/21/world/coronavirus-missing-deaths.html.

[17] E. Brown, A. Ba Tran y R. Thebault, «Excess U.S. deaths hit estimated 37,100 in pandemic's early days, far more than previously known,» The Washington Post, 2 May 2020. [En línea]. Available: https://www.washingtonpost.com/investigations/2020/05/02/excess-deaths-during-covid-19/?arc404=true.

[18] A. Voce, S. Clarke, C. Barr, N. McIntyre y P. Duncan, «Coronavirus excess deaths: UK has one of highest levels in Europe,» The Guardian, 29 May 2020. [En línea]. Available: https://www.theguardian.com/world/ng-interactive/2020/may/29/excess-deaths-uk-has-one-highest-levels-europe.

[19] D. Leon, V. M. Shkolnikov, L. Smeeth, P. Magnus, M. Pechholdová y C. I. Jarvis, «COVID-19: a need for real-time monitoring of weekly excess deaths,» 22 April 2020. [En línea]. Available: https://www.thelancet.com/pdfs/journals/lancet/PIIS0140-6736(20)30933-8.pdf.

[20] T. Cerna, «Ministro Mañalich descarta muertes por covid-19 no diagnosticadas: "La probabilidad es prácticamente nula",» Emol, 30 abril 2020. [En línea]. Available: https://www.emol.com/noticias/Nacional/2020/04/30/984716/Manalich-muertes-covido-no-diagnosticadas.html.

[21] M. C. Paredes, A. Faustino y C. Nazzal, «Tendencia de las defunciones ocurridas en mayores de 1 año según lugar de ocurrencia y su relación con características sociodemográficas, Chile 1997-2014,» *Rev Med Chile,* vol. 147, pp. 322-329, 2019.

[22] Registro Civil de Chile, «Portal de Datos del Servicio de Registro Civil e Identificación,» Registro Civil de Chile, mayo 2020. [En línea]. Available: https://estadisticas.sed.srcei.cl/defreg.

[23] M. Matus, «Exceso de muertes en COVID-19,» 2020. [En línea]. Available: https://github.com/marcelomatus/ExcesoChile.

[24] Ministerio de Ciencias, «Datos-COVID19,» [En línea]. Available: https://github.com/MinCiencia/Datos-COVID19.

[25] CNN Chile, «Casi no hay influenza ni virus sincicial: Cierre de colegios y menor uso de transporte público serían algunos factores,» CNN Chile, 22 May 2020. [En línea]. Available: https://www.cnnchile.com/coronavirus/casi-no-hay-influenza-ni-virus-sincicial_20200522/.

[26] Instituto Nacional de Estadísticas, «Estimación de personas extranjeras residentes en Chile,» 2019. [En línea].

[27] Y. Manríquez, «Ministro Mañalich descartó un aumento atípico en la mortalidad en Chile: "A veces se hacen matemáticas fáciles y conclusiones apresuradas",» ADN, 24 abril 2020. [En línea]. Available: https://www.adnradio.cl/nacional/2020/04/24/ministro-manalich-la-letalidad-del-covid-19-en-chile-es-una-de-las-mas-bajas-en-el-mundo-solo-superada-por-corea-del-sur.html.

[28] A. Jara, «Mañalich anuncia corrección en total de fallecidos con Covid-19: incorpora a 653 personas y eleva cifra a 2.190,» 7 mayo 2020. [En línea]. Available: https://www.latercera.com/nacional/noticia/manalich-anuncia-correccion-en-cifra-de-fallecidos-con-covid-19-incorpora-a-653-personas-y-eleva-cifra-total-a-2290/VR2N2AWOIZAGJHDTDRWA7AON3M/.

[29] A. Noufaily, D. G. Enki, P. Farrington, P. Garthwaite, N. Andrews y A. Charlett, «An Improved Algorithm for Outbreak Detection in Multiple Surveillance Systems,» *Statistics in Medicine,* vol. 32, nº 7, pp. 1206-1222, 2012.